\documentclass[10pt, conference]{IEEEtran}

\usepackage{bm,cite}
\usepackage{amsmath}
\usepackage{amsthm}
\usepackage{amsbsy}
\usepackage{epsfig}
\usepackage{latexsym}
\usepackage{amssymb}
\usepackage{wasysym}
\usepackage{placeins}  
\usepackage{color}
\usepackage{rotating}  
\usepackage[usenames,dvipsnames]{xcolor}
\usepackage{dsfont}
\usepackage{caption}
\usepackage{subcaption}
\usepackage{algorithm}
\usepackage{algorithmic}

\usepackage{paralist}

\usepackage{algorithm, algorithmic}

\DeclareMathAlphabet{\mathbit}{OML}{cmr}{bx}{it}
\DeclareMathAlphabet{\mathsf}{OT1}{cmss}{m}{n}
\DeclareMathAlphabet{\mathTXf}{OT1}{cmss}{bx}{it}

\DeclareMathOperator*{\argmax}{argmax}

\DeclareMathOperator{\PCSI}{PCSI}

\DeclareMathOperator{\Rate}{R}

\newcommand{\bG}{\mathbf{G}}

\newcommand{\bx}{\bm{x}}

\newcommand{\LB}{\left(}
\newcommand{\RB}{\right)}
\newcommand{\LSB}{\left[}
\newcommand{\RSB}{\right]}

\newcommand{\E}{{\mathbb{E}}}

\theoremstyle{remark}
\theoremstyle{assumption}

\makeatletter
\renewcommand{\maketag@@@}[1]{\hbox{\m@th\normalsize\normalfont#1}}%
\makeatother

\title{Decentralized Deep Scheduling for Interference Channels} 

\author{\\
\\
 Data Science Department, EURECOM}

 \author{
\IEEEauthorblockN{Paul de Kerret \IEEEauthorrefmark{3}, David Gesbert \IEEEauthorrefmark{3}, and Maurizio Filippone \IEEEauthorrefmark{4}} 

\IEEEauthorblockA{\IEEEauthorrefmark{3}
Communication Systems Department, EURECOM}

\IEEEauthorblockA{\IEEEauthorrefmark{4}Data Science Department, EURECOM}
}

\begin{document}

\maketitle
 

\begin{abstract}
In this paper\footnote{D. Gesbert and P. de Kerret are supported by the European Research Council under the European Union's Horizon 2020 research and innovation program (Agreement no. 670896).}, we study the problem of decentralized scheduling in Interference Channels (IC). In this setting, each Transmitter (TX) receives an arbitrary amount of feedback regarding the \emph{global multi-user} channel state based on which it decides whether to transmit or to stay silent \emph{without any form of communication with the other TXs}. While many methods have been proposed to tackle the problem of link scheduling in the presence of reliable Channel State Information (CSI), finding the optimally robust transmission strategy in the presence of arbitrary channel uncertainties at each TX has remained elusive for the past years. In this work, we recast the link scheduling problem as a decentralized classification problem and we propose the use of Collaborative Deep Neural Networks (C-DNNs) to solve this problem. After adequate training, the scheduling obtained using the C-DNNs flexibly adapts to the decentralized CSI configuration to outperform other scheduling algorithms.
\end{abstract} 

\IEEEpeerreviewmaketitle

\section{Introduction} \label{se:introduction}

\subsection{Decentralized Coordination in Wireless Networks}

Coordination between the Transmitters (TXs) in wireless network has received significant attention as a mean to improve Quality-of-Service (QoS) and spectral efficiency, through coordinated scheduling, interference reduction and alignment, joint beamforming, pilot coordination, and power control among many other possibilities. Coordination is often designed in a centralized setting whereby a computing node gathers all the necessary Channel State Information (CSI) from all TXs, computes utility maximizing decisions and forwards these decisions to the devices. In the realm of cellular networks, such a centralized implementation has been considered in the context of \emph{Cloud Radio Access Network (RAN)} or \emph{C-RAN} (See for example~\cite{Park2013}) supported by high-end all-optical backhaul architectures.

Yet considering the high cost and relative lack of flexibility of C-RAN deployments there is rising interest for decentralized forms of coordination. In emerging ultra-flexible and heterogeneous deployment scenarios featuring access points mounted on buses, drones \cite{Chen2017}, or those allowing backhaul-less Device-to-Device communications, decentralized control becomes a highly desirable feature as it allows for cheaper, faster, and more flexible coordination schemes. 

However, decentralized coordination between the TXs also comes with its own challenges. While the centralized implementation allows to perfectly share CSI from all TXs (so-called logically centralized), in contrast the devices in a decentralized setting must cope with their own local uncertainties regarding the global CSI in order to make a transmission decision, which calls for innovative algorithm designs.  
 
Indeed, in the distributed CSI configuration, the TXs aim at cooperating on the basis of different locally available information, which can be recast as decentralized Team Decision (TD) problem \cite{Radner1962}. The key challenge in TD stems from the fact that each decision maker (here, device) is limited by the local uncertainties (noise) affecting its view of the global CSI, making it difficult to predict the behavior of other devices with which it seeks to coordinate its actions. In prior work, attempts to derive noise-robust policies are reported \cite{Li2015b}, yet always relying on scenario specific policy models and heuristics, making the implementation non-generic. In \cite{dekerret2016_SPAWC,Zhang2017} more generic approaches are proposed using discrete sets, making the approaches non-scalable.
 
In this work, we show how \emph{Collaborative} Deep Neural Networks (DNNs) under a suitable learning strategy allow for a generic approach to robust decentralized coordination. The principles of our approach are exemplified through the case of link scheduling in wireless interference channels~\cite{Gjendemsjo2008}. 

We first give the reader a quick background on classical DNNs before moving to the newer decentralized scenario.

\subsection{Supervised Deep Learning: A (Very) Short Overview}\label{se:Supervised_Learning}
In supervised deep learning, we aim at \emph{learning} a mapping~$f$ between $\bx_i$ and $f(\bx_i)$ from a training data set~$(\bx_i,f(\bx_i))_{i=1}^n$. In this work, we model the function using \emph{Deep Neural Network (DNN)} such that the function is restricted to be obtained from the output of a multiple-layer feed-forward DNN. A DNN consists of multiple layers where the $j$th layer contains $n_j$ nodes. The output of each node is then obtained from a linear combination of the outputs of the previous layer followed by the application of a so-called \emph{activation} function which introduces the required non-linearity. Thus, a DNN is obtained from the composition of non linear functions, where each function is a linear combination of activation functions. 

Specifically, let us denote by $y_i^j$ the output of the $i$th node of layer~$j$ and by $\Phi$ the activation function. The output of node~$i$ of layer~$j$ is then given by
\begin{equation}
y_i^j=\Phi\LB \sum_{i=1}^{n_j} \theta_i^{j-1} y_i^{j-1}\RB 
\label{eq:intro_1}
\end{equation}
where $\theta_i^{j}$ for all values of $i$ and $j$ form the parameters of the DNN that need to be \emph{trained}. Clearly, the last layer has a number of node corresponding to the output space while the first layer corresponds to the input space. Activation functions are chosen so as to reach the desired accuracy in the training at the fastest rate. Currently, the most widely used activation function is the so-called \emph{ReLu} function given by
\begin{equation}
\mathrm{ReLu}(z)= \max(z, 0).
\label{eq:intro_2}
\end{equation}
One important advantage of the ReLu function is that its derivative is either $0$ or $1$ and hence easily implemented. DNNs have been known for many years but were notably difficult to train until recent breakthroughs both in terms of hardware and in terms of algorithms, which made possible computationally efficient training of DNNs \cite{Lecun2012,Lecun2015}. 

\subsection{DNN Literature Overview}
DNN has been applied to many different scenarios, achieving striking performances and successes \cite{Lecun2015,Mnih2015} yet mostly related to computer vision or speech processing. 

Turning to wireless communication, the application of the new deep learning tools is only at its infancy, although recently gaining a lot of momentum.  DNNs have been used in some cases to reproduce (approximate) known algorithms. The advantage of this approach is that the demanding computations are then done during the training of the DNNs. Once the DNNs coefficients are obtained, the use of the DNN requires only very simple computations, thus allowing for quasi-real time processing. This approach is studied in \cite{Lei2017} for caching and in \cite{Sun2017} for resource allocation in interference channels.

In \cite{Clancy2007} a framework to incorporate machine learning in cognitive radio is described while its application to coding is discussed in \cite{Cammerer2017,Gruber2017}. In \cite{Samuel2017}, deep learning is used for detection to reduce complexity while it is used in \cite{Farsad2017} to make up for the absence of channel model while performing detection in molecular communication. In \cite{Awan2017}, learning is applied to the determination of the optimal cell-load in wireless networks by deriving and exploiting mathematical properties of the problem considered. 

In \cite{Oshea2017,Dorner2017}, the use of deep learning to design the physical layer is discussed and it is in particular shown how the TX, the channel, and the RX can been seen as a single DNN and trained as an autoencoder, which is a model combining encoding and decoding that are learned jointly.
\subsection{Main Contributions}
In contrast to the previous literature on deep learning for communication systems, in this work we propose the use of a novel collaborative type of DNNs (C-DNNs) with the specific goal to enable robust transmission schemes in a decentralized multi-device setting with uncertainties. To the best of our knowledge, such an application has not been considered before. In this context we bring the following contributions:
 \begin{itemize}
\item We consider C-DNNs consisting of multiple parallel DNNs that are used to derive transmission policies in a coordinated wireless network scenario.
\item We propose a suitable training strategy for the C-DNNs allowing for robustness with respect to an arbitrary configuration of channel feedback uncertainties existing at each device (decision maker).
\item We apply the above principle to the example of link scheduling in wireless interference channels with arbitrary noisy channel feedback at each TX. We show how the scheduling obtained using the trained C-DNNs outperforms other scheduling schemes in terms of network sum throughput.
\end{itemize}

\section{System Setting}\label{se:system_model}

\subsection{Transmission Scenario} 
We consider a $K$-user Interference Channel (IC) consisting of $K$ single-antenna Transmitters (TXs) and $K$ single-antenna Receivers (RXs) with RX~$i$ being only served by TX~$i$. The gain of the wireless channel between TX~$j$ and RX~$i$ is denoted by $G_{i,j}$ and all the gains are put together to form the \emph{channel gain matrix}~$\bG\in \mathbb{R}^{K\times K}$ where 
\begin{align}
\{\bG\}_{i,j}\triangleq G_{i,j}.
\label{eq:SM_1}
\end{align}
The channel gain matrix can be generated according to any probability density function $p_{\bG}$ \emph{known to all TXs} as it is a long term information that can be estimated and shared among all TXs . We study link scheduling which means that each TX may decide between transmitting with a fixed maximum power level~$P$ or staying idle for one channel realization. This is akin to a binary power control problem whereby the power level $P_j$ at TX~$j$ can take its values in $\{0,P\}$. Note that the proposed approach is easily extendeable to additional discrete power levels. We assume for ease of notation that all TXs have maximum power constraint~$P$ and that the RXs undergo a unit variance Gaussian noise. 

We further assume that the data symbols transmitted are distributed as i.i.d. Gaussian and that each RX treats interference as noise such that the instantaneous sum rate is given by \cite{Cover2006}
\begin{equation}
R(P_1,\ldots, P_k)\!= \!\sum_{k=1}^K\log_2\!\left(\!1+\frac{G_{k,k} P_k}{1+\sum_{\ell=1,\ell\neq k}G_{k,\ell} P_{\ell}}\!\right). 
\label{eq:SM_2}
\end{equation}  
Our common \emph{welfare} objective in this work will be the expected sum rate. With perfect knowledge of the gain matrix, maximizing the expected sum rate comes down to maximizing the sum rate for each individual channel realization such that the optimal power control function~$p_1^{\PCSI},\ldots,p_K^{\PCSI}$ can be obtained from:
\begin{equation}
\begin{aligned}
&(p_1^{\PCSI}(\bG),\ldots,p_K^{\PCSI}(\bG)) \!=\!\!\!\!\!\!\!\!\argmax_{(P_1,\ldots,P_K)\in \{0,P\}^K}\!\!\!\!\!\!\!\Rate\LB \bG , P_1,\ldots,P_K\RB 
\end{aligned}
\label{eq:SM_PCSI}
\end{equation}
Binary forms of power control were shown to be rate optimal in the case of $2$-user IC with perfect CSI at all TXs and near optimal with more users~\cite{Gjendemsjo2008}. Importantly, perfect link scheduling at every TX requires (logically) centralized and perfect knowledge of the matrix gain $\bG$ above.

\subsection{Distributed Channel State Information} 
Due to imperfect CSI feedback and limited CSI sharing links across TXs, each TX is now assumed endowed with its own imperfect estimate of the current channel state. Specifically, TX~$j$ obtains the estimate~$\hat{\bG}^{(j)} \in \mathbb{C}^{K\times K}$ of the channel gain matrix and chooses its transmission power~$P_j$ as a function of~$\hat{\bG}^{(j)}$, \emph{without any form of information exchange with the other TXs}. 

This distributed CSI model is very general as it allows for any joint distribution $p_{\bG,\hat{\bG}^{(1)},\ldots,\hat{\bG}^{(K)}}$. The estimates at the different TXs can for example be correlated, and in the limiting case where the estimates at all TXs are exactly equal, the (logically) \emph{centralized} CSI configuration is recovered.

\subsection{Team Decision Problem} 
Based on the locally available channel state information~$\hat{\bG}^{(j)}$, TX~$j$ choses its binary power control~$P_j$. Yet, the optimal instantaneous choice of $P_j$ would normally depend on the power control decisions at the other TXs, which are unknown. Consequently, we need to introduce the power control \emph{strategies} denoted by~$p_1,\ldots,p_K$ and given by
\begin{equation}
\begin{array}{cccc}
p_j:  & \mathbb{R}^{K\times K} & \rightarrow & \{0,P\}\\
       &   \hat{\bG}^{(j)}  & \mapsto & p_j(\hat{\bG}^{(j)} )
\end{array}
\label{eq:SM_3}
\end{equation}
Specifically, the TXs aim at jointly maximizing the expected sum rate, such that the TD problem formulated is given by
\begin{equation}
(p_1^{\star},\ldots,p_K^{\star})\!=\!\!\!\!\argmax_{(p_1,\ldots,p_K)\in \mathcal{P}}\!\!\mathbb{E}\!\LSB \!\Rate\LB\bG, p_1(\hat{\bG}^{(1)}),\ldots,p_K(\hat{\bG}^{(K)})\!\RB \!\RSB
\label{eq:SM_4}
\end{equation}
where the expectation is carried out across all random variables, i.e., according to $p_{\bG,\hat{\bG}^{(1)},\ldots,\hat{\bG}^{(K)}}$ and $\mathcal{P}$ is defined by
\begin{equation}
\mathcal{P}\triangleq\left \{(p_1,\ldots, p_K)|p_j : \mathbb{R}^{K\times K}\rightarrow \{0,P\}\right \}.
\label{eq:SM_5}
\end{equation} 
Optimization problem \eqref{eq:SM_4} is a functional optimization problem, as the optimization is done over the power control functions~$p_j$, hence making this problem particularly challenging to solve with classical optimization tools. 

To evaluate the C-DNN scheduling which will be introduced in Section~\ref{se:CDNN}, we start by presenting simpler approaches that do not take into account the intricate distributed CSI configuration and can hence serve as references to evaluate how much theC-DNNs are able to adapt to the distributed CSI configuration.

\subsubsection{Naive Link Scheduling}\label{se:naive}
In the distributed CSI configuration, the so-called \emph{naive} link scheduling strategy $p_j^{\mathrm{naive}}$ is obtained when TX~$j$ (erroneously) assumes that (i) all TXs have received the same estimate $\hat{\bG}^{(j)}$ and that (ii) this estimate is noiseless. Consequently, its decision is obtained by solving the optimization problem with perfect CSI given in \eqref{eq:SM_PCSI} only using $\hat{\bG}^{(j)}$ instead of $\bG$. Considering for exposition TX~$1$, the naive scheduling strategy~$p_1^{\mathrm{naive}}$ is obtained from:
\begin{equation}
\begin{aligned}
&(p_1^{\mathrm{naive}}(\hat{\bG}^{(1)}),q_2^{\star}(\hat{\bG}^{(1)}),\ldots,q_K^{\star}(\hat{\bG}^{(1)}))\\
&=\argmax_{(P_1,\ldots,P_K)\in \{0,P\}^K} \Rate\LB\hat{\bG}^{(1)}, P_1,P_2,\ldots,P_K\RB 
\end{aligned}
\label{eq:SM_7}
\end{equation}
where we have denoted the ``optimal" decision at TX~$j$ with $j\neq 1$ by $q_j^{\star}$ to emphasize the fact that this power allocation strategy is only an auxiliary variable: It will never be used because it simply corresponds to an erroneous guessing at TX~$1$.

\subsubsection{Locally Robust Link Scheduling}\label{se:locally_robust}
The Locally Robust (LR) link scheduling corresponds to the intermediate step between the \emph{naive} link scheduling and the \emph{decentralized robust} link scheduling. This means that the imperfect nature of the estimate~$\hat{\bG}^{(j)}$ is considered, but not the decentralized nature of the problem. As both $\bG$ and $\hat{\bG}^{(j)}$ arise in the expectation, it is not possible to simplify the optimization as it was the case for naive scheduling. Hence, at TX~$1$, the LR scheduling is obtained from:
\begin{equation}
\begin{aligned}
&(p_1^{\mathrm{LR}},r_2^{\star},\ldots,r_K^{\star})\\
&\!=\!\!\argmax_{(p_1,p_2,\ldots,p_K)\in \mathcal{P}}\!\!\!\!\E \!\LSB\!\Rate\!\LB\!\bG\!, p_1(\hat{\bG}^{(1)})\!,p_2(\hat{\bG}^{(1)}),\!\ldots,p_K(\hat{\bG}^{(1)})\!\RB\!\RSB 
\end{aligned}
\label{eq:SM_8}
\end{equation} 
For arbitrary CSI configurations and channel distribution, the LR scheduling is not known. Consequently, for the sake of comparison, we will use the C-DNNs link scheduling applied \emph{locally} at each TX using the locally available estimate~$\hat{\bG}^{(j)}$ only. This will allow us to evaluate how C-DNNs scheduling is able to adapt to the distributed CSI structure.

\section{Collaborative Deep Neural Network}\label{se:CDNN}

\subsection{Principle}

The team decision problem formulated in \eqref{eq:SM_4} is a challenging problem. It requires optimizing over multiple functions depending on different input variables such that there are two main technical difficulties (i) the functional aspect and (ii) the decentralized nature of the information. Furthermore, the optimal robust strategy has to be flexible enough to adapt to any CSI configuration: When all TXs have perfect CSI, the optimal link scheduling strategy consists in simply using instantaneously the best link scheduling, i.e., applying the naive strategy given in \eqref{eq:SM_7}. In contrast, when no CSI is available anywhere, some form of orthogonal scheduling will come into play. For all intermediate configurations, the optimal robust link scheduling strategy should navigate in between and exploit whatever local and asymmetric information is available. We will now show how C-DNNs can overcome these two problems to efficiently provide a robust scheduling function.

In the proposed C-DNNs framework, the transmission strategy at TX~$j$ is parametrized by a DNN denoted by $p^{\bm{\theta}_j}$ taking as input the multi-user estimate~$\hat{\bG}^{(j)}$ and returning as output the binary decision~$P_j$. Intuitively, the power control strategy $p_j$ is replaced by the DNN power control~$p^{\bm{\theta}_j}$. Consequently, the complete strategy space is reduced to the space that can be parametrized by the DNN. To avoid introducing any sub-optimality, it is necessary that the DNN space is large enough such that it contains the optimal power control strategy $p^{\star}$ or approximates it asymptotically well. This will be ensured by choosing a DNN with enough coefficients, i.e., large/deep enough. In addition, it is also necessary that the coefficients~$\bm{\theta}_j$ can be \emph{trained} efficiently to reach their optimal values.

With this parametrization, the original TD problem is now approximated as
\begin{equation}
(\bm{\theta}_1^{\star},\ldots,\bm{\theta}_K^{\star})=\!\argmax_{\bm{\theta}_1,\ldots,\bm{\theta}_K} \mathbb{E}\LSB \!\Rate\!\LB\! \bG, p^{\bm{\theta}_1}_1\!(\hat{\bG}^{(1)})\!,\ldots,p^{\bm{\theta}_K}_K\!(\hat{\bG}^{(K)})\!\RB\!\RSB\!.
\label{eq:DDNN_1}
\end{equation}
The key aspect of \eqref{eq:DDNN_1} comes from the fact that although optimization problem \eqref{eq:DDNN_1} is not a supervised learning problem, the coefficients~$\bm{\theta}_j$ can be optimized using learning methods briefly introduced in Section~\ref{se:Supervised_Learning}. Indeed, the instantaneous objective is the rate function defined in \eqref{eq:SM_2} which is differentiable such that gradient-based optimization methods can be used.

As all the probability density functions are assumed to be known, the training data set, denoted by $\mathcal{S}^{\mathrm{train}}_n$ can be obtained from $n$~Monte-Carlo realizations of the channel and the channel estimates generated according to $p_{\bG,\hat{\bG}^{(1)},\ldots,\hat{\bG}^{(K)}}$ such that
\begin{equation}
\mathcal{S}_n^{\mathrm{train}}\triangleq \left \{\LB\bG_i,\hat{\bG}_i^{(1)},\ldots,\hat{\bG}_i^{(K)}\RB\big| i=1,\ldots, n\right \}.
\end{equation}
Generating this training set can be seen as approximating the expectation in \eqref{eq:DDNN_1} using Monte-Carlo realizations to yield
\begin{equation}
\begin{aligned}
&(\bm{\theta}_1^{\star},\ldots,\bm{\theta}_K^{\star})\\
&\approx \argmax_{\bm{\theta}_1,\ldots,\bm{\theta}_K}\frac{1}{n}\sum_{i=1}^n \Rate\LB \bG_i, p^{\bm{\theta}_1}_1(\hat{\bG}^{(1)}_i),\ldots,p^{\bm{\theta}_K}_K(\hat{\bG}^{(K)}_i)\RB
\end{aligned}
\label{eq:DDNN_2}
\end{equation}
where the approximation becomes exact as the number of samples~$n$ increases to infinity. 

It is important to note that the training of the link scheduling functions is done \emph{jointly} at all TXs, while the \emph{application} of the link scheduling functions is decentralized at each TX. Note that the joint training is possible as it depends only on the statistics which are known to all TXs. Formulation~\eqref{eq:DDNN_2} allows for the use of deep learning tools implemented in high levels packages such as TensorFlow to train the C-DNNs. Yet, note that the training of DNNs is known to be sensitive to DNN parameters (number of training steps, learning rate, ...).
\begin{figure}[htp!] 
\centering
\includegraphics[width=1\columnwidth]{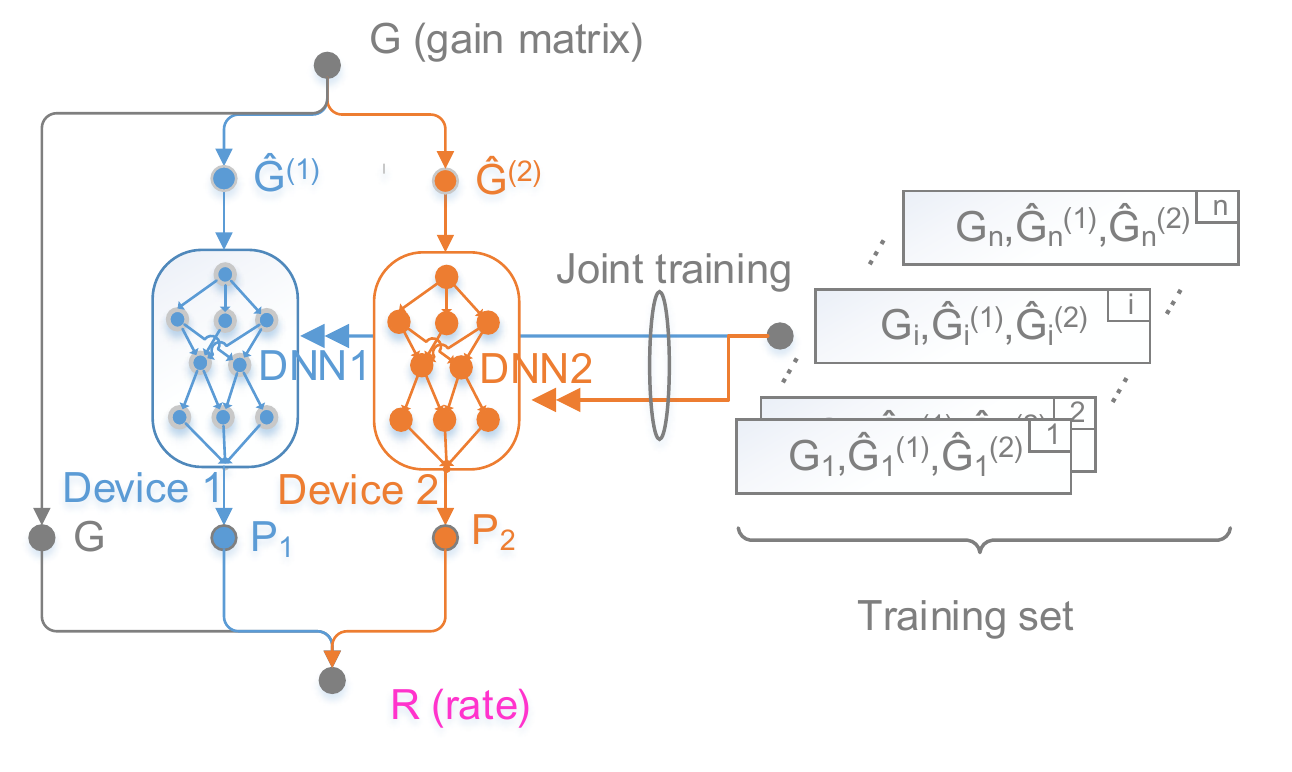}
\caption{Illustration of collaborative-DNNs (C-DNNs) applied to link scheduling in the $2$-user IC. Key aspects are the joint training using the common knowledge of the distribution and the decentralized application of the scheduling algorithm.} 
\label{ICC2018_Decentralized_DNN}
\end{figure}  

\subsection{Initialization using Naive Scheduling}
Using the right initialization is key to an efficient training \cite{Sutskever2013}. We have chosen to initialize our DNN at TX~$j$ with the coefficient~$\bm{\theta}_j$ that best approximates the naive link scheduling~$p^{\mathrm{naive}}_j$ in \eqref{eq:SM_7}. We start by generating a training codebook from Monte-Carlo realizations:
\begin{equation}
\mathcal{T}_n^{\mathrm{naive}}\triangleq \{(\bG_i,p^{\mathrm{naive}}_j(\bG_i))\}_{i=1}^n
\end{equation}
Using the training dataset~$\mathcal{T}_n^{\mathrm{naive}}$, the coefficients~$\bm{\theta}_j$ can be trained using any conventional supervised learning algorithm. 

Using this initialization comes with a second important advantage: It allows us to evaluate how the DNNs are able to efficiently approximate the naive scheduling~$p^{\mathrm{naive}}_j$ wich is this time a conventional supervised learning problem. This supervised learning constitutes hence a first test to select the right coefficients for the C-DNNs.


\section{Experiments}\label{se:Experiments}
In our experiments, we use a TensorFlow implementation of a $3$-layer DNN with fully connected layers comprising $30$ neurons each and using the ReLu activation function defined in~\eqref{eq:intro_2}. We have used $n=30000$ Monte-Carlo realizations with batch size of $5000$ realizations. We furthermore use a drop-out probability equal to~$0.5$ at each node to avoid overfitting \cite{Srivastava2014}. Finally, we run the Adam gradient based optimizer $10000$ times with a learning rate of $0.001$. 

We further consider Rayleigh fading such that the channel gains are distributed as i.i.d. Chi-square random variables. To model the distributed CSI configuration, we also consider for simplicity an additive Gaussian model such that the estimate at TX~$j$ is given by
\begin{equation}
\hat{\bG}^{(j)}\triangleq \bar{\bm{\Sigma}}^{(j)}\odot\bG+ \bm{\Sigma}^{(j)}\odot\bm{\Delta}^{(j)}
\label{eq:experiments_1}
\end{equation}
where $\odot$ is the element-wise (Hadamard) product, $\bm{\Delta}^{(j)}$ contains i.i.d. Chi-square random variables and $\bm{\Sigma}^{(j)}$ is the matrix containing the variance of the CSI noise at TX~$j$ while $\bar{\bm{\Sigma}}^{(j)}$ is defined such that
\begin{equation}
\{\bar{\bm{\Sigma}}^{(j)}\}_{i,k}\triangleq \sqrt{1-\{\bm{\Sigma}^{(j)}\}_{i,k}^2},\qquad \forall i,k\in \{1,\ldots,K\}.
\label{eq:experiments_2}
\end{equation}

We will compare the trained C-DNNs with the following schemes:
\begin{itemize}
\item \emph{Perfect CSI scheduling}: This corresponds to the optimal link scheduling with perfect CSI instantaneously at all TXs, and is hence clearly an (a priori) loose outerbound.
\item \emph{Always-on Scheduling}: All TXs are always active.
\item \emph{TDMA}: Only one of the TX is active for all channel realizations.
\item \emph{Naive scheduling}: See Section~\ref{se:naive}.
\item \emph{Locally Robust (LR) scheduling}: See Section~\ref{se:locally_robust}.
\end{itemize} 
\subsection{Preliminaries: The Centralized Case}
As a preliminary result, we start by considering a centralized setting with $K=2$~TXs so as to verify the performance of the LR DNN scheduling. It hence holds~$\hat{\bG}^{(1)}=\hat{\bG}^{(2)}$. We further use the following covariance matrix
\begin{equation}
\bm{\Sigma}^{(1)}=\begin{bmatrix}0&\sigma\\ \sigma& 1 \end{bmatrix}
\label{eq:experiments_3}
\end{equation}
with $\sigma\in [0,1]$. It is important to understand that a $1$ means no information relative to the true channel but only noise, while $0$ means perfect knowledge of the coefficient. Finally, to give more learning opportunities to the DNN, we also consider that the variance of the cross-channel from TX~$2$ to RX~$1$ is attenuated by $0.25$.

\begin{figure}[htp!] 
\centering
\includegraphics[width=1\columnwidth]{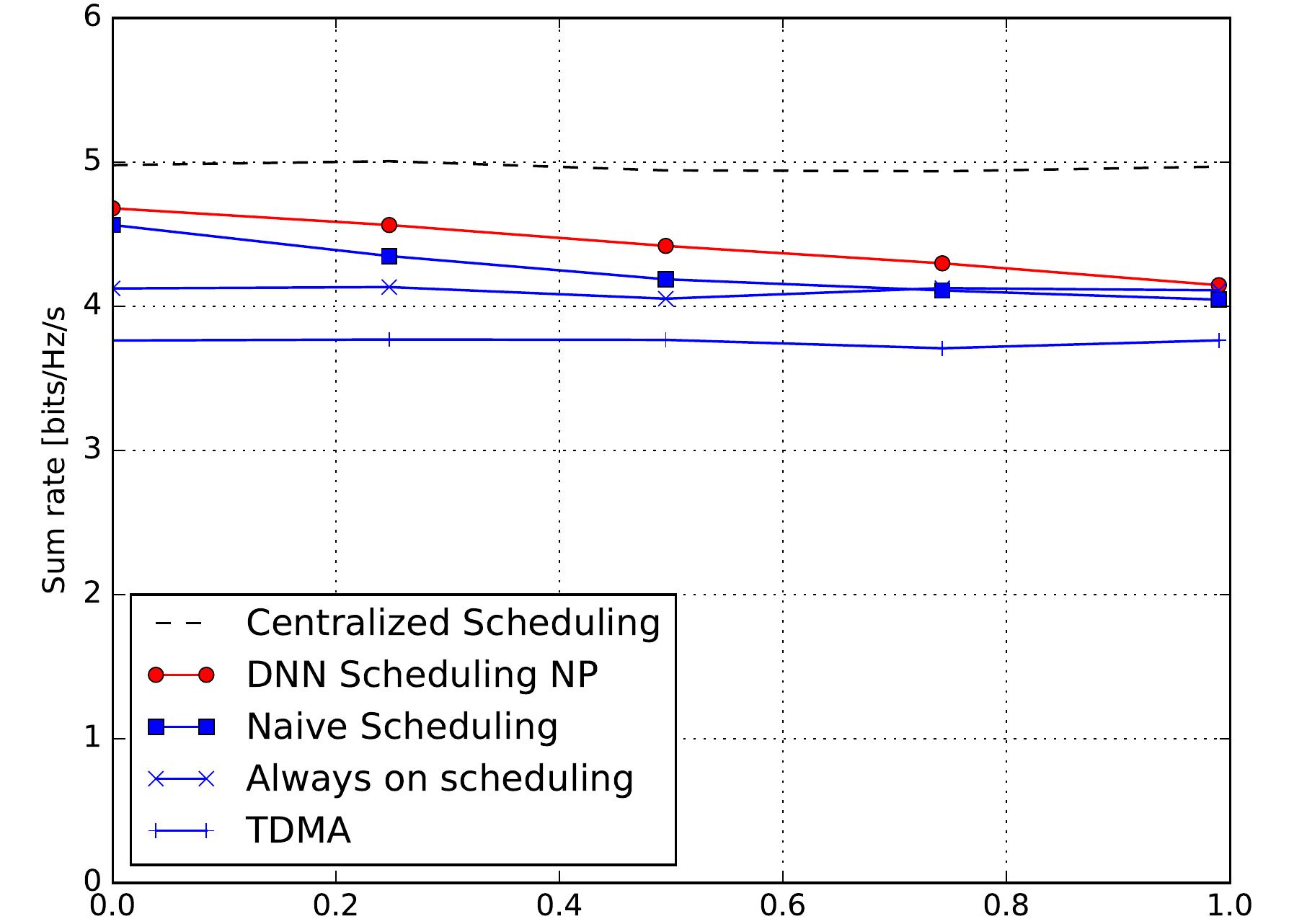}
\caption{} 
\label{ICC2018_2users_rate_centralized}
\caption{Expected sum rate as a function of the CSI quality parameter~$\sigma$.}
\end{figure}

In Fig.~\ref{ICC2018_2users_rate_centralized}, we show the expected sum rate achieved as a function of the $\sigma$ parameter in \eqref{eq:experiments_3}. It can be seen how the C-DNN scheduling outperforms naive scheduling. Yet, the gain is reduced due to the fact that in the centralized CSI configuration, there is always \emph{perfect} coordination.

\begin{figure*}[htp!]
    \centering
    \begin{subfigure}[t]{0.5\textwidth}
        \centering
        \includegraphics[width=1\columnwidth]{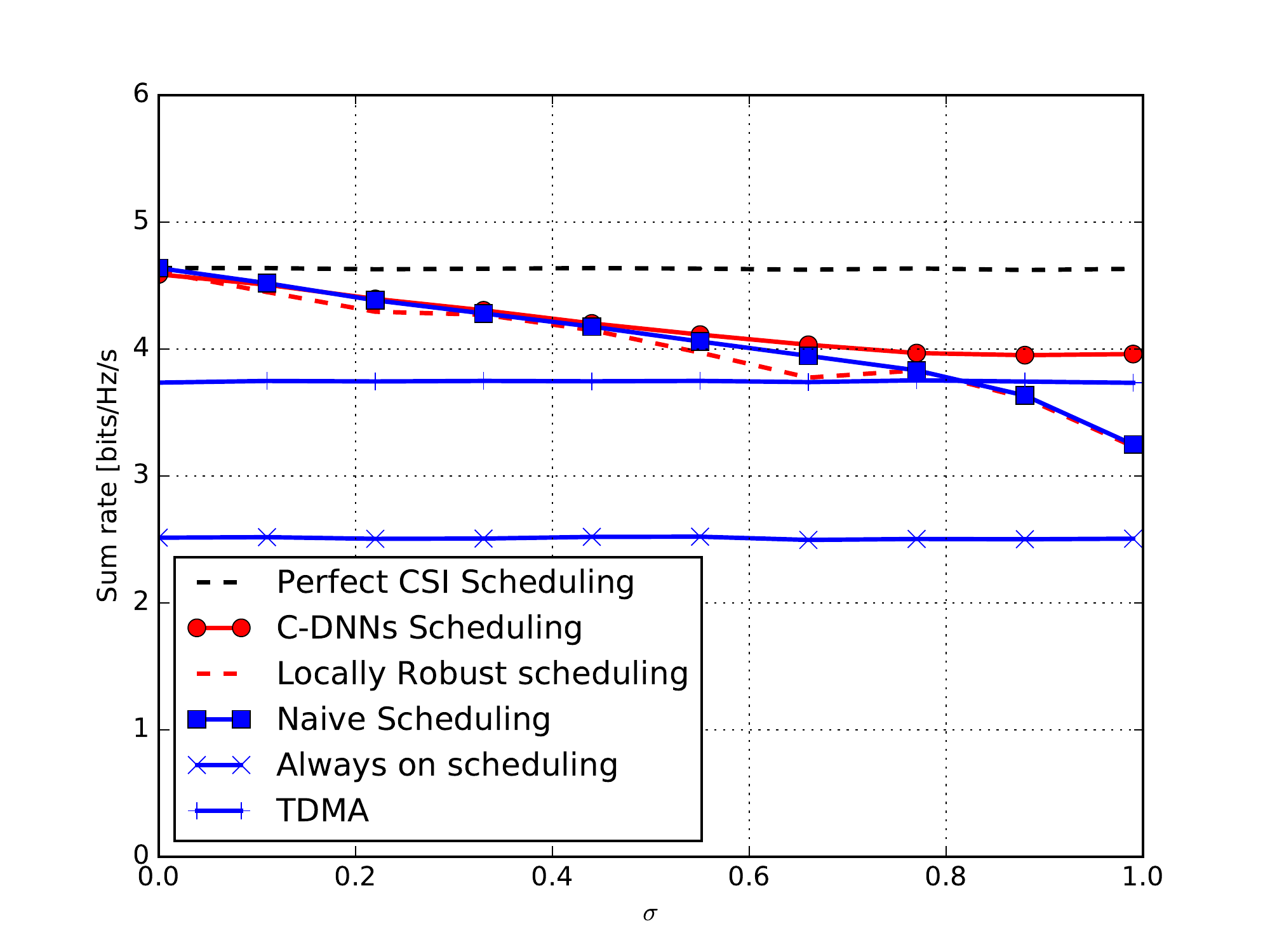}
        \caption{Expected sum rate as a function of the CSI quality parameter~$\sigma$.}
				\label{ICC2018_2users_rate_scenario1}
    \end{subfigure}%
    ~ 
    \begin{subfigure}[t]{0.5\textwidth}
        \centering
        \includegraphics[width=1\columnwidth]{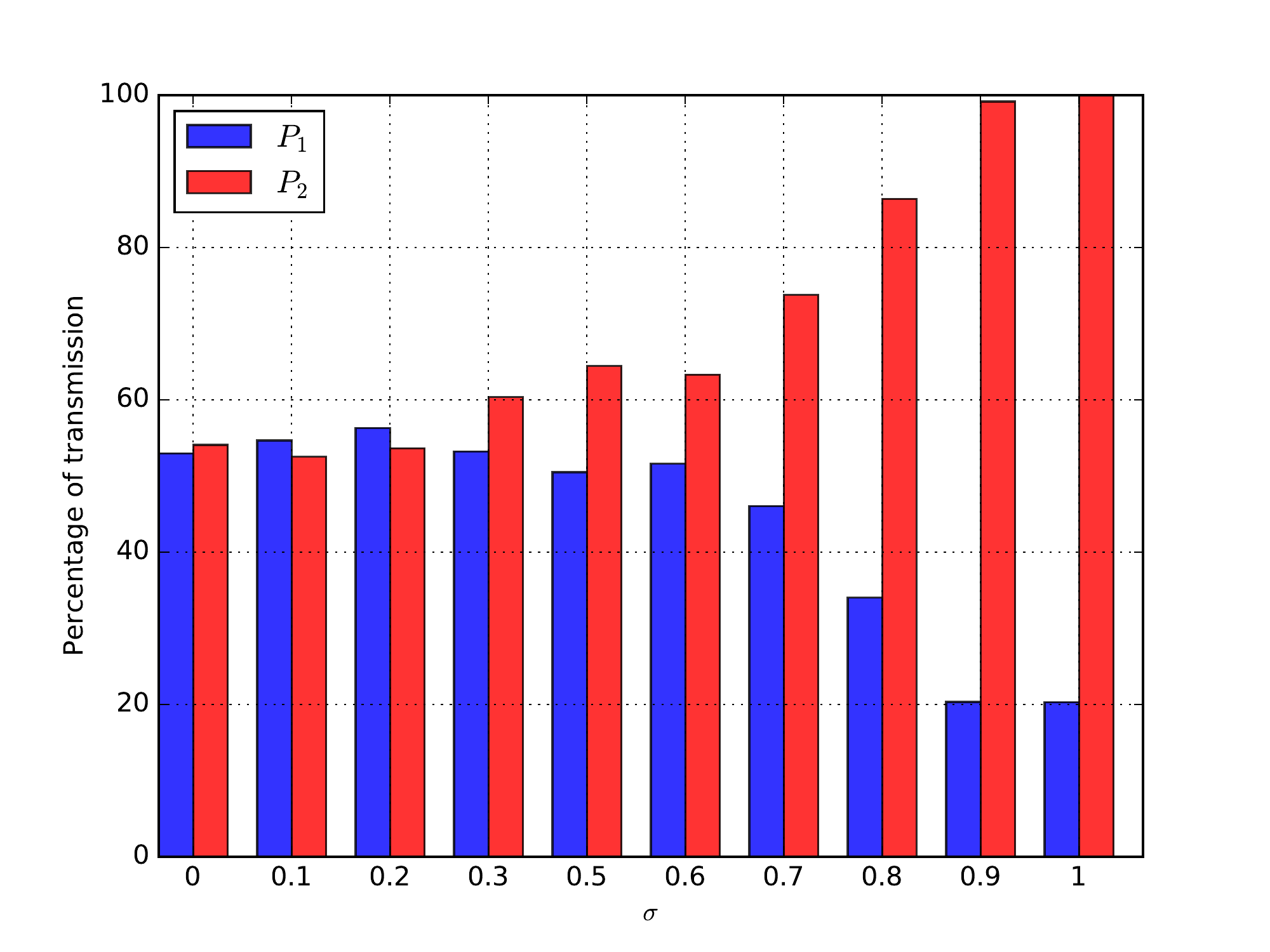}
        \caption{Percentage of transmission as a function of the CSI quality parameter~$\sigma$. As the CSI quality at TX~$1$ degrades, the strategy at TX~$2$ becomes more deterministic with TX~$1$ adapting to this configuration.}
    \end{subfigure}
    \caption{Simulations results in a $2$-user IC with TX~$1$ having perfect CSI and TX~$2$ having no CSI.}
				\label{ICC2018_2users_power_scenario1}
\end{figure*}
\begin{figure*}[htp!]
    \centering
    \begin{subfigure}[t]{0.5\textwidth}
        \centering
        \includegraphics[width=1\columnwidth]{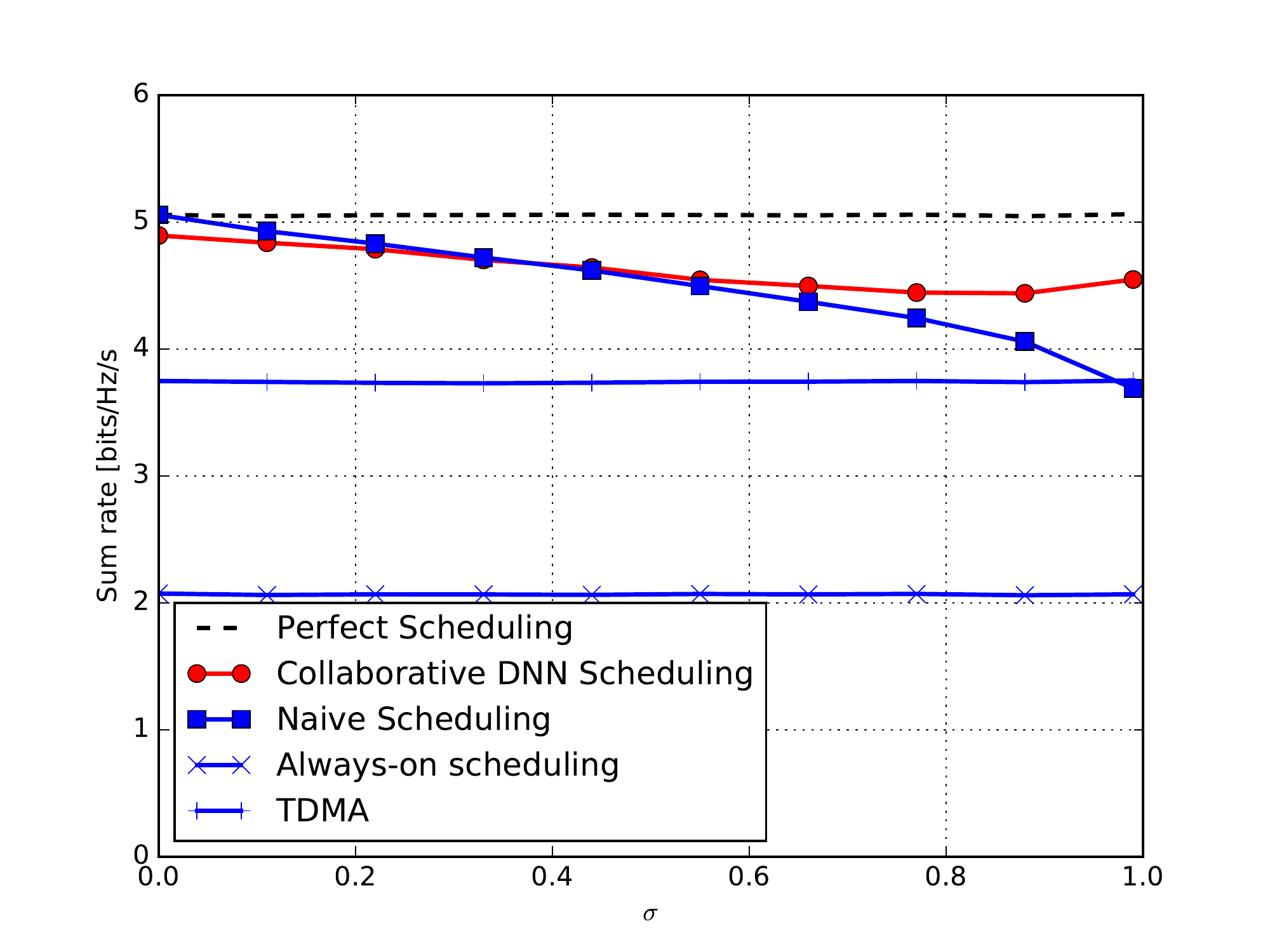}
        \caption{Expected sum rate as a function of the CSI quality parameter~$\sigma$.}
				\label{ICC2018_3users_rate_scenario1}
    \end{subfigure}%
    ~ 
    \begin{subfigure}[t]{0.5\textwidth}
        \centering
        \includegraphics[width=1\columnwidth]{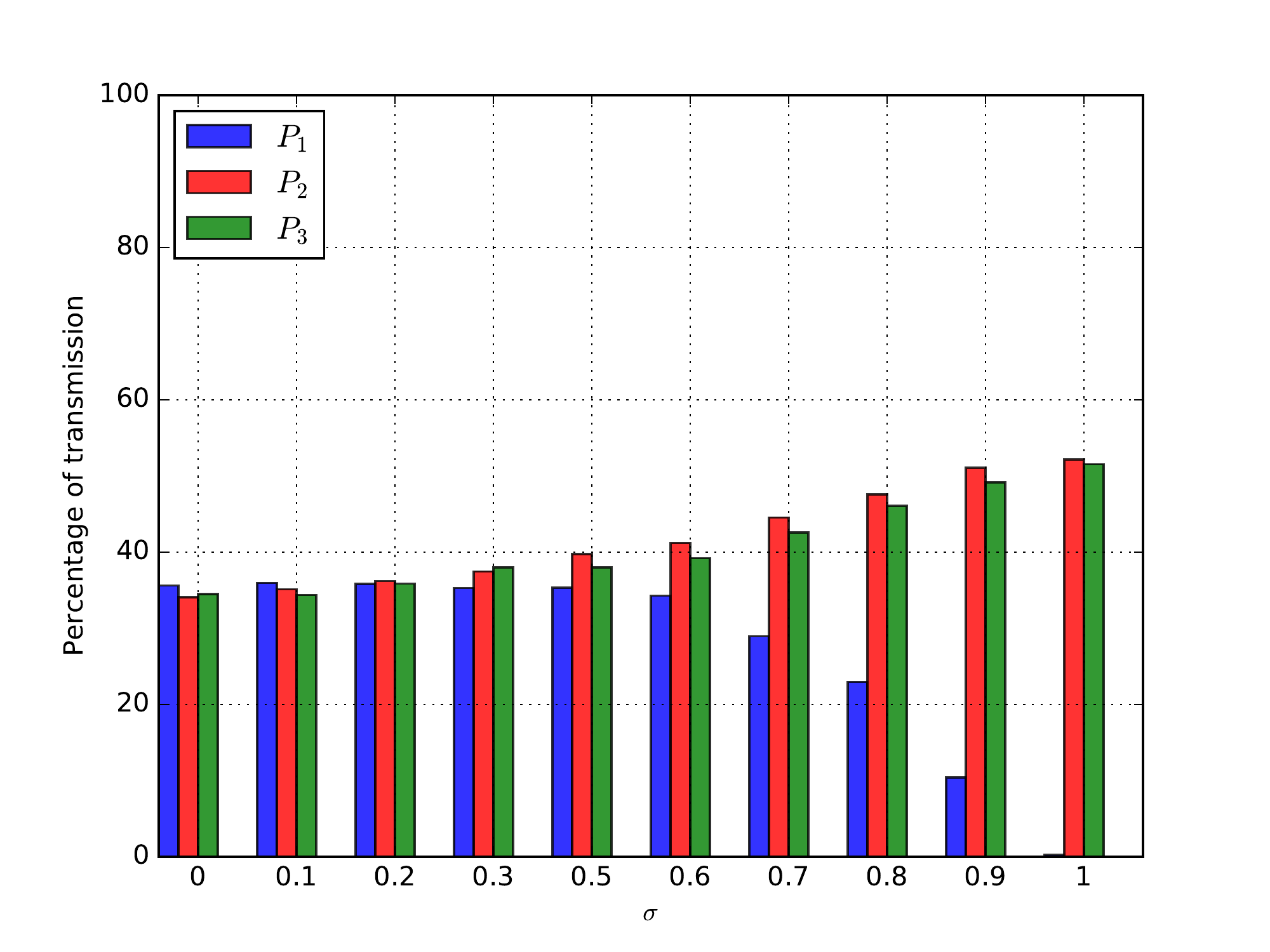}
        \caption{Percentage of transmission as a function of the CSI quality parameter~$\sigma$. With accurate CSI at all TXs, naive scheduling is used. When the CSI quality at TX~$3$ degrades, it starts transmitting less, before being completely shut down.}
				\label{ICC2018_3users_power_scenario1}
    \end{subfigure}
    \caption{Simulations results in a $3$-user IC with TX~$1$ having partial CSI and TX~$2$ and TX~$3$ having no CSI.}
\end{figure*}

\subsection{Two-User Interference Channel}
We now reconsider a $2$-users distributed CSI configuration with the CSI noise covariance matrices given by
\begin{equation}
\bm{\Sigma}^{(1)}=\begin{bmatrix}\sigma&\sigma\\ \sigma & \sigma \end{bmatrix},\quad \bm{\Sigma}^{(2)}= \begin{bmatrix}0&0\\ 0& 0 \end{bmatrix}.
\label{eq:experiments_3}
\end{equation}
In Fig.~\ref{ICC2018_2users_rate_scenario1}, the LR scheduling does not improve significantly from the naive scheduling. In contrast, C-DNNs scheduling goes as expected from naive scheduling with near perfect CSI to a solution outperforming TDMA when one TX is fully uninformed. Looking at the percentage of transmission for each TX in~Fig.~\ref{ICC2018_2users_power_scenario1} when $\sigma=1$, the uninformed TX transmits all the time while the informed TX adapts to this transmission using its perfect estimate, which fits with the intuition.

\subsection{Three-User Interference Channel}

We now turn to a $3$-user IC with the following CSI noise covariance matrices:
\begin{equation}
\bm{\Sigma}^{(1)}= \sigma \mathbf{1}_{3\times 3},\quad \bm{\Sigma}^{(1)}=\bm{\Sigma}^{(2)}= \mathbf{0}_{3\times 3}.
\label{eq:experiments_4}
\end{equation}
Intuitively, two TXs are perfectly informed, and one TX has uniform intermediate CSI. In Fig.~\ref{ICC2018_3users_rate_scenario1}, when $\sigma$ is near zero, all TXs have practically perfect CSI and C-DNNs scheduling is slightly outperformed by naive scheduling. This can be understood as the need for a larger training set, i.e., for more computing power. Furthermore, the performance of the C-DNNs scheduling seems to slightly increase with $\sigma$ when $\sigma$ is large, which is either due to problems of convergence to local optima or to the size of the training set. Yet, its analysis is outside the scope of this work.

\section{Conclusion}	
In this work, we have shown how DNNs could be use collaboratively to obtain an efficient robust solution to a challenging decentralized link scheduling problem. The proposed robust solution outperforms previously known methods and is able to adapt to any distribution of channel and CSI configuration. Improving the training of the DNNs and finding the optimal C-DNNs architecture are very interesting topics for future works.

\bibliographystyle{IEEEtran}
\bibliography{Literature}
\end{document}